\documentclass{elsart3}

\usepackage{graphics}
\usepackage{amssymb}
\begin{document}

\begin{frontmatter}


% \ead{flouquet}

\title{Phase diagram of heavy fermion systems}


\author[label1,label4]{J.~Flouquet\corauthref{cor1}},
\ead{flouquet@cea.fr}
\author[label2]{Y.~Haga},
\author[label3]{P.~Haen},
\author[label1]{D.~Braithwaite},
\author[label1]{G.~Knebel},
\author[label1]{S.~Raymond},
\author[label2]{S.~Kambe}

\corauth[cor1]{Corresponding author. Tel.: +33-438-785423; FAX
+33-438-785098}

\address[label1]{D\'{e}partement de Recherche Fondamentale sur la Mati\`{e}re Condens\'{e}e,
SPSMS, CEA Grenoble, 38054 Grenoble, France}
\address[label2]{Advanced Science Research Center JAERI, Tokai, Ibaraki 319-1195, Japan}
\address[label3]{CRTBT-CNRS, BP 166X, 38042 Grenoble-C\'{e}dex, France}
\address[label4]{Department of Physical Science, Graduate School of Engineering Science, Osaka University, Japan}

\begin{abstract}

The Meccano of heavy fermion systems is shown on different cases
going from anomalous monochalcogenides to cerium intermetallic
compounds with special focus on the ideal case of the
CeRu$_2$Si$_2$ series. Discussion is made in the frame of the
interplay between valence, electronic structure (Fermi surface),
and magnetism. The nice tools given by the temperature, the
pressure, and the magnetic field allow to explore different ground
states as well as the slow downhill ``race'' before reaching a
Fermi liquid finish line at very low temperature. Experimentally,
the Gr\"{u}neisen parameter i.e.~the ratio of the thermal
expansion by the specific heat is a coloured magic number; its
temperature, pressure, and magnetic field dependence is a deep
disclosure of competing hierarchies and the conversion of this
adaptive matter to external responses.

\end{abstract}

\end{frontmatter}

% main text

Heavy fermion physics are approached by pedestrian looking of
basic questions on the electronic occupancy $n_{4f}$ of the 4$f$
shell and the magnetic properties. The monochalcogenide cases of
Sm and Tm compounds are revisited with recent data obtained on SmS
\cite{Haga03,Barla03}. For the cerium intermetallic compounds,
special attention is given close to the so called magnetic quantum
criticality i.e.~near the critical electronic density $\rho_c$ or
pressure $P_c$ where drastic changes occur on the ground state.
Pressure and magnetic field are elegant tool to modify the weight
or even the sign of the magnetic interactions by comparison to the
strong local fluctuations inherent to Kondo lattices. Emphasis is
given on the results on the CeRu$_2$Si$_2$ familly since the pure
compound i.e.~with vanishing disorder allows to scan through
different temperature and field regimes \cite{Flouquet02}.

\section{Monochalcogenides of Sm and Tm: conduction, valence, and magnetism}

\begin{figure}
 \scalebox{.37}{\includegraphics*{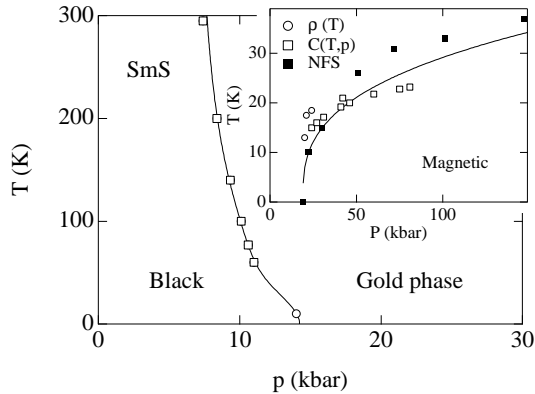}}
 \caption{\label{figure1}Phase diagram of SmS with the first order line between the black and the gold phase.
  Insert: Zoom on the metallic gold phase and the discovery of magnetism by three different techniques:
  specific heat (C), transport ($\rho$) and nuclear forward scattering (NFS). }
\end{figure}

In the monochalcogenides of anomalous rare earths (RE) like Sm or
Tm the equilibrium between two valence states (Sm$^{2+}$/Sm$^{3+}$
or Tm$^{2+}$/Tm$^{3+}$) governs the release of a 5$d$ electron
\cite{Wachter94}. Let us start with our new guided tour of the
canonical example of SmS. The full line $T_{B-G}$ of the ($T-P$)
phase diagram (see Fig.\ref{figure1}) represents the first order
isostructural transition between the black (B) insulating (I)
Sm$^{2+}$ phase and the so called gold (G) phase where the Sm ions
jumps to an intermediate valence state $v \approx 2.7$. In this
gold phase, smoothly increasing the pressure must lead to recover
a trivalent Sm$^{3+}$; it is a Kramers ion, long range magnetic
order must appear above this threshold $P=P_c$.

Furthermore a sound experimental fact is that up to $P_\Delta =
20$~kbar in the low pressure gold phase a many body insulating
phase is smoothly formed on cooling. Whatever the sample quality
and the pressure conditions, $P_\Delta$ is a fixed pressure above
which the sample is clearly metallic i.e.~its resistivity $\rho$
decreases on cooling after reaching a maximum $\rho_{max}$ at
$T_m$ \cite{Holtzberg81,Lapierre81,Konzykowski81,Ohashi98}. The
recent combination of macroscopic measurements by transport and ac
specific heat and of the microscopic probe of the hyperfine field
on the Sm nuclei by nuclear forward scattering (NFS) using ESRF
facility demonstrates that long range antiferromagnetic (AF) order
appears at $P_c \sim P_\Delta$. As the pressure variation of $T_m$
now identified as the N\'{e}el temperature $T_N$ is huge, $P_c$ is
presumably a first order point associated with a lattice parameter
discontinuity i.e.~a tiny valence jump at $P_v = P_c$ towards the
trivalent state.

\begin{figure}
 \begin{center}
 \scalebox{.55}{\includegraphics*{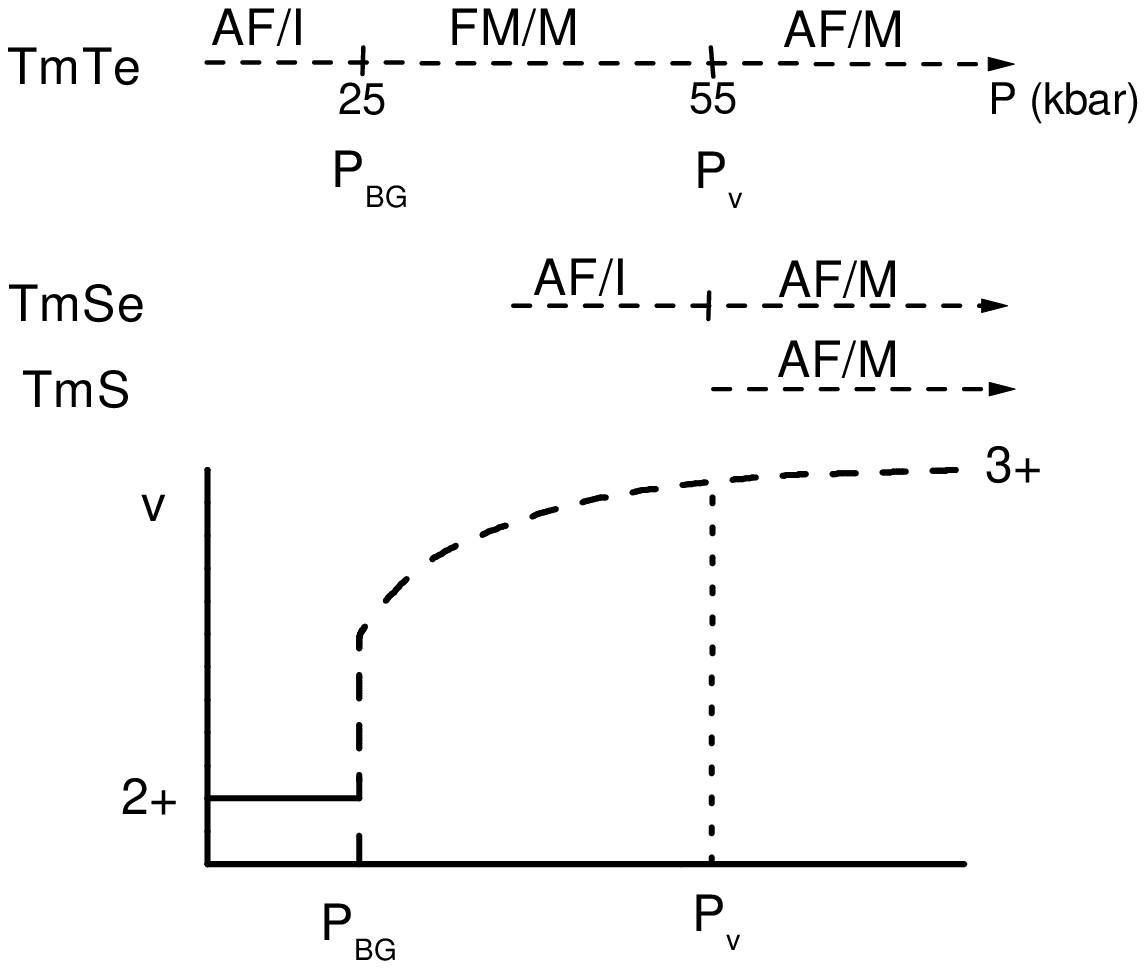}}
 \caption{\label{figure2}Pressure cascade of ground states in Tm chalcogenides: $v$ is the rare earth valence. }
 \end{center}
\end{figure}

By contrast to the Sm case where the valence fluctuations occur
between the Sm$^{2+}$ state with a zero angular momentum ($J=0$)
and Sm$^{3+}$ with a $J=5/2$ angular momentum, for the Tm ions the
mixing is between two finite angular momentum states, respectively
$J=7/2$ (Tm$^{2+}$) and $J=6$ (Tm$^{3+}$). Whatever the pressure
i.e.~the valence and even the electronic conduction, the ground
state is magnetic. However, the similarity with SmS is a link
between magnetic properties and 4$f$ localisation. For TmTe as
shown on Fig. \ref{figure2}, the $P$ sequence of ground states is
AF/I up to $P_{BG}$, ferromagnetic (F) and metallic (M) up to
$P_v$ and finally the entrance to a AF metallic state
\cite{Link97,Mignot01}; the rapid collapse of $T_{Curie}$ just
below $P_v$ and the fast increase of $T_N$ just above $P_v$
suggests a first order transition. For TmSe, already in the gold
phase at $P=0$, the switch at $P_v$ occurs between an AF
insulating phase $P< P_v - \epsilon$ and another AF metallic state
above $P_v$ \cite{Ribault80}. TmS itself appears already at $P=0$
as a quiet quasitrivalent metallic state, in principle the ideal
Kondo lattice \cite{Holtzberg85}.

\section{Cerium heavy fermion systems: valence and magnetism}

As pointed out already, the cerium heavy fermion problem can be
summarized by two major ($T,P$) phase diagrams. The wellknown
$T_{\alpha,\gamma}$ first order transition of the cerium metal and
the up to date magnetic phase diagram $T_N$ or $T_{Curie}$ versus
pressure where the ordering temperature may collapse at $P_c$
\cite{Kambe00}. For the cerium metal, the treatment of the
magnetism can be crude as all characteristic temperatures
$T_{\alpha,\gamma}$ are high by comparison to any hypothetical
$T_N$ or any fine structure as the crystal field splitting
$C_{CF}$ of the trivalent $\gamma$ phase. At $P=0$,
$T_{\alpha,\gamma}(0)$ is already positive, finite and near 100~K
and the critical end points $T_{cr}=700$~K, $P_{cr}=22$~kbar are
large. For typical cerium heavy fermion compounds $P_c$ is lower
or comparable to $P_{cr}$ and of course $T_{\alpha,\gamma}(P=0)$
and $T_{cr} (P_{cr})$ will be negative. However, the 4$f$ electron
will feel the distance from $P_{cr}$ and cause the crossing
through a characteristic pressure $P_v$ where the local
susceptibility loses the anisotropy given by its reaction to the
crystal field (that corresponds to the well known condition $k_B
T_K$ (Kondo energy) greater than $C_{CF}$). Above $P_v$, the 4$f$
electron stays rigidly linked to the full degenerancy of the
$J=5/2$ angular momentum. At $P_v$, the sensitivity of the 4$f$
electron to the crystal field has vanished; $P_v$ is
characteristic of a valence crossover. The relative location of
$P_c$ and the hidden parameter $P_v$ is one of the key issue. The
idea that singular properties is underpinned by a quantum critical
point i.e.~ pressure is one of the trends of strongly correlated
electronic systems (see references \cite{Bourdarot03} for the
hidden order of URu$_2$Si$_2$ or \cite{Chiao03} for the
metamagnetic quantum criticality in Sr$_3$Ru$_2$O$_7$ and
\cite{Nakatsujii03} for the conversion of a Kondo gas to heavy
electron Kondo liquid).

Recently special focus has been given on systems just on the
paramagnetic verge of $P_c$ i.e.~at $P_c+\epsilon$
\cite{Schroeder00}. The Non Fermi liquid (NFL) label refers to the
difficulty to reach the standard Fermi liquid regime where usual
laws are found, even for non-interacting Fermi gas (specific heat
$C=\gamma T$ linear in temperature $T$, susceptibility in $T^2$,
scattering rate proportional to $T^2$); the NFL label guarantee
``vintage'' describes a very large temperature crossover before
reaching the Fermi liquid graal or a new state of matter.

\begin{figure}
\begin{center}
 \scalebox{.45}{\includegraphics*{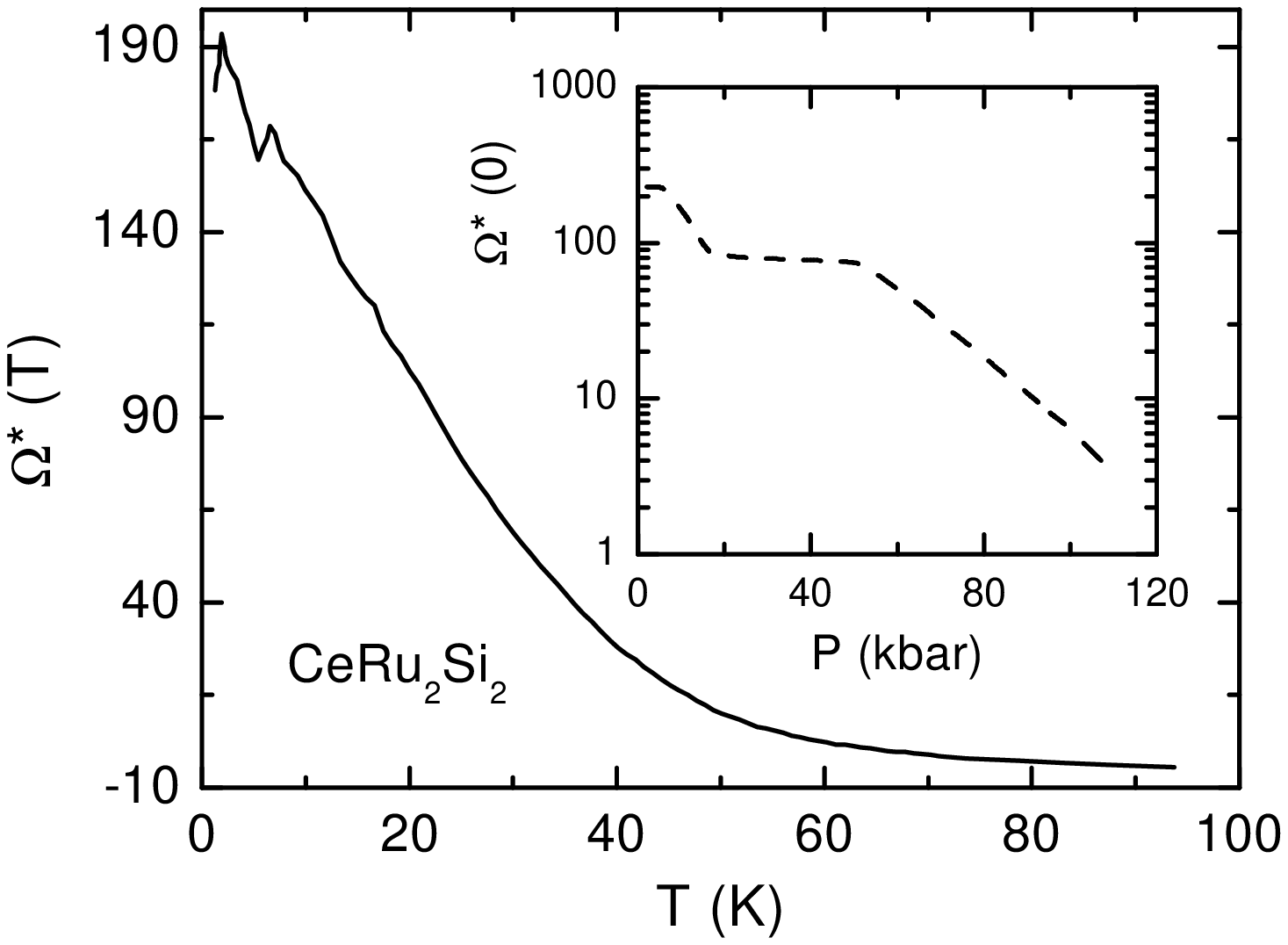}}
 \caption{\label{figure3}Temperature variation of the Gr\"{u}neisen parameter $\Omega^\star (T)$ of CeRu$_2$Si$_2$.
 Insert: pressure dependence of $\Omega^\star (0)$, the extrapolated zero temperature limit.  }
\end{center}
\end{figure}

To clarify the complexity of this strongly correlated electronic
system, an old fashioned approach via thermodynamic languages will
be carried out \cite{Benoit81,Flouquet82,Kambe97,Zhu03}. In the
same spirit as the Clapeyron and Ehrenfest relations on the $P$,
$T$ feedback for first or second order transitions, the
Gr\"{u}neisen parameter
\begin{equation}
\Omega^\star (T) = \frac{\alpha}{C}\frac{V_0}{\kappa}
\end{equation}
(where $\alpha$, $V_0$ and $\kappa$ are respectively the volume
thermal expansion, the molar volume and the isothermal
compressibility) is a unique experimental gauge of interfering
hierarchies. It will be reduced to a constant $\Omega^\star (0)$
only if the free energy $F$ can be expressed by a unique parameter
$T^\star$ i.e.~
\begin{equation}
F (T) = T\Phi (\frac{T}{T^\star}) \quad \rm{with} \quad
\Omega^\star (0)= - \frac{\partial \log T^\star}{\partial \log V}
\end{equation}
Figure \ref{figure3} represents the temperature variation of the
Gr\"{u}neisen parameter of CeRu$_2$Si$_2$ at $P=0$ i.e.~roughly 3
kbar above the critical pressure $P_c=-3$~kbar. The two singular
points are: (i) the huge extrapolated value of $\Omega^\star (0) =
+190$ and (ii) the slow entrance in a simple regime ($T \approx
1$~K) where $\alpha$ and $C$ reach their proportionality
\cite{Lacerda89,Holtmeier89}. The pressure dependence of
$\Omega_0^\star$ is shown in the insert of fig.\ref{figure3}. At
low pressure close to $P_c$, $\Omega^\star (0)$ is weakly pressure
dependent changing from 190 to 220 for a negative shift of $P =
2$~kbar \cite{Lacerda89} while $\Omega_0^\star$ drops to 80 at
$P_v$ \cite{Holtmeier89}. Other values of $\Omega_P^\star (0)$ can
be listed crudely for other heavy fermion system. In a double
logarithmic representation of the volume dependence of the $A$
coefficient of the well known $AT^2$ Fermi liquid resistivity
assuming the usual Kadowaki-Woods rule $A \propto \gamma ^2$,
$\Omega^\star (0)$ is directly linked to the slope of the curve
\cite{Flouquet03}.

\section{Magnetic field switch from AF correlation to F interactions}

The compound CeRu$_2$Si$_2$ is a quasi-ideal Kondo lattice with a
nice ordered tetragonal lattice, a robust Ising character of the
local susceptibility, the realization of a clean limit condition
i.e.~a mean free path higher than 1000 \AA\ far larger than an
hypothetical superconducting coherence length with temperature
$T_s = 200$~mK and good enough to obtain the complete
determination of the Fermi surface (FS) at least above $P_c$ (see
references in \cite{Flouquet02}).

\begin{figure}
\begin{center}
 \scalebox{.4}{\includegraphics*{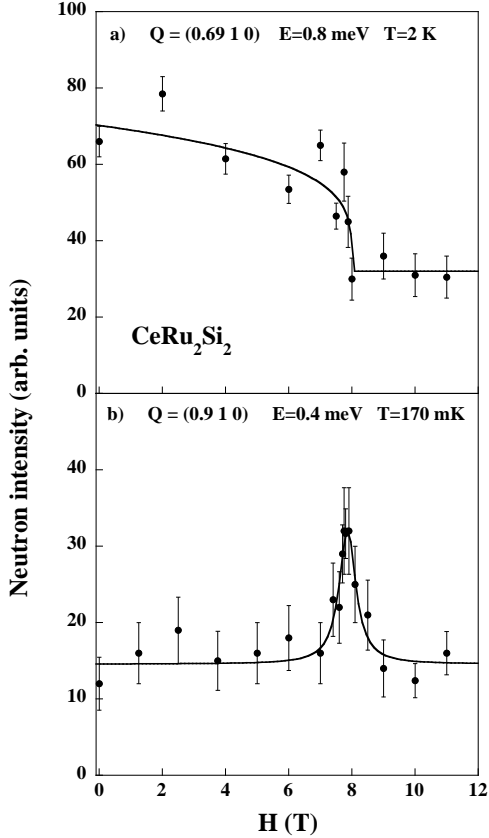}}
 \caption{\label{figure4}(a) Field variation of the inelastic response at the energy transfer $E=0.8$~meV measured
 at the AF wave vector (0.69, 1, 0); the AF contribution vanishes at
 $H_m$ while the local one survives.
  (b) Appearance of low energy ferromagnetic fluctuation ($E=0.4$~meV) on crossing
 $H_m$. }
\end{center}
\end{figure}

The application of a magnetic field $H$ along the easy $c$ axis
leads to a switch through a drastic continuous crossover at
$H=H_m$ from a low field paramagnetic phase (Pa) dominated by AF
correlations at a finite wavevector $k_0$ to a highly polarized
state (PP) dominated by the low wavevector $q$ excitation. The
microscopic vision of CeRu$_2$Si$_2$ was nicely given by inelastic
neutron scattering experiments, (i) with at $H=0$, the balance
between the local response and the intersite coupling, (ii) with,
under magnetic field, the continuous spreading of the AF response
with the same characteristic glumly energy $\omega_{AF} \sim
1.6$~meV up to $H_m$ \cite{Raymond99} and the observation just in
the vicinity of $H_m$ of an inelastic ferromagnetic signal at far
lower energy transfer $\omega_F \sim 0.4$~meV than $\omega_{AF}$
(see Fig.\ref{figure4}) \cite{Sato01,Raymond03}.

A skilful mechanism characteristic of the Kondo lattice
CeRu$_2$Si$_2$ controles the ``ping-pong'' between the magnetic
interactions. Quite remarkably, the field and temperature response
is well reproduced by a simple form of the thermodynamic as the
entropy.
\begin{equation}
S=S(\frac{T}{T^\star},\frac{H}{H_m})
\end{equation}
with equal Gr\"{u}neisen parameters $\Omega_{T^\star} =
\Omega_{H_m} = -190$. With this hypothesis one can predict the
field variation of $\gamma_H$ knowing the field variation of the
linear thermal expansion $\alpha_v=a_H T$. A satisfactory
agreement is found between this scaling derivation of $\gamma_{H}$
and the bare determination either by specific heat or by
magnetization\cite{Lacerda89,Paulsen90}. Physically, the adaptive
formation of the many body Kondo lattice bands is characterized by
a pseudogap . The Ising spin character of the bare local magnetic
ion plays a capital role in the sharpness of this electronic
substructure and correlatively for the pseudo-metamagnetism
\cite{Aoki98}. The pseudogap ingredient is a key input in any
theory which needs to respect the scaling behavior
\cite{Ohkawa98}. Reminiscent of the $\alpha-\gamma$ collapse of
the cerium metal, the associated spectacular lattice softening
($\approx 30$\%) is again in agreement with the push pull between
magnetic, electronic and lattice instabilities \cite{Flouquet02}
and the underpinned vicinity of a quantum critical end point.

\section{To be or not to be? Speculations on superconductivity}

After a lecture describing the ``beauty'' CeRu$_2$Si$_2$ to
quantum criticality, the first attack is to produce the proofs
that CeRu$_2$Si$_2$ is close to a magnetic instability. Looking
back to the primitive results one decade ago on the CeRu$_2$Si$_2$
series either sweeping electronic density (doping or pressure) or
magnetic field, a NFL increase of $C/T$ on cooling in a given
temperature window justifies the proximity to $P_c$
\cite{Fisher91}. Furthermore, this variation is not drastically
different from that popularized in the canonical example of
CeCu$_{5.9}$Au$_{0.1}$ \cite{Loehneysen94} after a rescaling of
the temperature by some phenomenological parameter describing the
proximity to $P_c$ and also differences in physical ingredients.
As on approaching $P_c$, no drastic pressure dependence of
$\Omega^\star (0)$ occurs. Collapsing to $P_c$ on the paramagnetic
side does not lead to an ``earthquake'' even if of course
$T^\star$ will collapse when $T \to 0$. The deviation from any
singular scaling law of the imaginary part of the dynamical
susceptibility $\chi''T^\alpha \propto \Phi(\frac{\omega}{T})$
becomes also quite difficult to point out. The yet unanswered
dilemma is if a new state  of matter occurs right at $P_c$.

The obsessive focus just at $P_c$ or on the paramagnetic (Pa) side
($P>P_c$) leads to neglect the disappearance of long range
magnetism. As many tunings were realized by doping i.e.~at the
cost of an increase of disorder, $P_c$ appears as a second order
quantum phase transition. Comparison with the previous
chalcogenide cases, recent NMR experiments on CeIn$_3$ close to
$P_c - \epsilon$ with evidence of AF and Pa magnetic phase
separation \cite{Kawasaki03} as well as drastic Fermi surface
changes at $P_c$ in CeRh$_2$Si$_2$ \cite{Araki01} point out that
at the limit of a pure lattice at $P_c$ the transition may be
first order. Today one key question is to determine precisely the
Fermi surface on both sides of $P_c$ knowing the two limit
conditions of a localized 4$f$ picture at low pressure ($P\ll
P_c$) and an itinerant 4$f$ picture at high pressure ($P\gg P_c$).
Progress will imply a deep interdisciplinary collaboration between
experimentalists and theoreticians eager to track carefully the
$P$ and $H$ variations including the modification of the Brillouin
zone and the differentiation between spin up and spin down
effective mass carriers.

For unconventional superconductivity, the nature of the phase
transition at $P_c$ is not critical as proved by the occurrence of
superconductivity in CeIn$_3$ and CeRh$_2$Si$_2$. There are
examples like CePd$_2$Si$_2$, CeIn$_3$ and CeRh$_2$Si$_2$ where
the superconducting pocket is stocked to $P_c \sim P_v$; there are
cases like CeCu$_2$Si$_2$ \cite{Bellarbi84}, CeCu$_2$Ge$_2$
\cite{Jaccard92} and CeNi$_2$Ge$_2$ \cite{Braithwaite00} where two
superconducting domains are centered on distinct pressures $P_c$
and $P_v$. We will test soon if superconductivity appears on the
verge of $P_v \approx 20$~kbar in CeRu$_2$Si$_2$; near $P_c =
-3$~kbar, the Ising character of the spin allows only longitudinal
magnetic fluctuations which does not favor the observation of $d$
wave pairing \cite{Monthoux02,Lonzarich03} while near $P_v$ this
superconductivity can be promoted by the emergence of transverse
magnetic modes as well as valence fluctuations
\cite{Onishi,Holmes03}.

It is time to concentrate on the complete bundle of curves on each
side of $P_c$ in CeRu$_2$Si$_2$ as well as in the other cerium
tetragonal lattices and Yb homologues  following our Dresden
colleagues \cite{Paschen03}.

\ack{J.~F.~wants to thank the Lorentz center for its invitation to
the workshop ``Non Fermi liquid behavior and quantum phase
transitions'' (May 2003) and FERLIN ESF support during four
years.}

% \appendix

% \section{}
% \label{}

\end{document}